# Fluctuation Theorem for Hamiltonian Systems -
# Le Chatelier's Principle

Denis J. Evans[1], Debra J. Searles[2], Emil Mittag[1]

[1] Research School of Chemistry, Australian National University, Canberra, ACT 2601, Australia

[2] School of Science, Griffith University, Brisbane, Qld 4111, Australia

For thermostatted dissipative systems the Fluctuation Theorem gives an analytical expression for the ratio of probabilities that the time averaged entropy production in a finite system observed for a finite time, takes on a specified value compared to the negative of that value. Hitherto it had been thought that the presence of some thermostatting mechanism was an essential component of any system which satisfies a Fluctuation Theorem. In the present paper we show that a Fluctuation Theorem can be derived for purely Hamiltonian systems, with or without applied dissipative fields.

The Fluctuation Theorem [1-3] (FT) gives a general formula for the logarithm of the probability ratio that in a thermostatted dissipative system, the time averaged entropy production $\overline{\Sigma}_t$ takes a value, A, to minus the value, -A.

$$\Pr(\overline{\Sigma}_t / k_B = A) \big/ \Pr(\overline{\Sigma}_t / k_B = -A) = \exp[At] \qquad (1)$$

From this equation it is obvious that as the averaging time or system size increases it becomes exponentially likely that the entropy production will be positive. The Fluctuation Theorem is important for at least three reasons: firstly it gives an expression for the probability that in a finite system observed for a finite time, the Second Law will be violated; secondly it gives one of the very few exact expressions for nonequilibrium steady states - even far from equilibrium; thirdly it can be derived using some of the standard results of the mathematical theory of dynamical systems theory [3].

The theorem was initially proposed [1] for nonequilibrium steady states that are thermostatted in such a way that the total energy of the system was constant. Subsequently, it was shown by Gallavotti and Cohen [3] that the theorem could be proved for sufficiently chaotic isoenergetic nonequilibrium systems using the SRB measure. For transient trajectory segments that start at t = 0 from an equilibrium ensemble of initial phase space vectors, $\mathbf{\Gamma}(0)$ and evolve in time under the influence of a reversible deterministic thermostat and an applied dissipative field, towards a unique nonequilibrium steady state a Transient Fluctuation Theorem can be derived using the Liouville measure [2, 4]. It has also been shown that the theorem is valid for a wide class of stochastic nonequilibrium systems [5].

It has recently be shown [4] that if initial phases are sampled from a known N-particle phase space distribution function, $f(\mathbf{\Gamma},0)$, and if we define a general dissipation function, $\Omega(\mathbf{\Gamma})$, by

$$
\begin{aligned}
t\overline{\Omega}_t &\equiv \int_0^t ds\, \Omega(\mathbf{\Gamma}(s)) \\
&= \ln\!\left(\frac{f(\mathbf{\Gamma}(0),0)}{f(\mathbf{\Gamma}(t),0)}\right) - \int_0^t \Lambda(\mathbf{\Gamma}(s))ds
\end{aligned}
\qquad (2)
$$

where $\Lambda(\mathbf{\Gamma}) \equiv \partial\!\big/\!\partial\mathbf{\Gamma} \bullet \dot{\mathbf{\Gamma}}$ is the phase space compression factor, then one can derive a Transient Fluctuation Theorem,

$$\frac{\Pr(\overline{\Omega}_t = A)}{\Pr(\overline{\Omega}_t = -A)} = \exp[At]. \tag{3}$$

Thermostats lead to nonzero expressions for the phase space compression factor. Equation (2) is consistent with all known deterministic Transient Fluctuation Theorems covering a wide variety of initial ensemble types and thermostatting mechanisms [4].

This relationship (3) has been tested using computer simulations for a range of nonequilibrium steady state systems in which the phase space contraction is non-zero [2, 4, 5b, 6]. It predicts that the dissipation function has a definite sign that is consistent with the Second Law of Thermodynamics. In the present paper we point out that equation (2) can be applied to purely conservative systems where there is no phase space contraction, $\Lambda(t) \equiv 0$. We consider two cases in detail: the adiabatic (unthermostatted) response of a system to a dissipative applied field, and secondly the free relaxation of density inhomogeneities in a system to which no dissipative fields or thermostats are applied. These examples are of general interest since the fine grained Gibbs entropy, $S_G(t) \equiv \int d\Gamma\, f(\Gamma(t)) \ln[f(\Gamma(t))]$ is a constant of the motion in both systems.

Consider a system of N interacting particles subject to a colour field $F_c$. The total Hamiltonian is $H(\boldsymbol{\Gamma}) = H_0(\boldsymbol{\Gamma}) + F_c \sum_{i=1}^{N} c_i y_i$, where $c_i = (-1)^i$ and $H_0(\boldsymbol{\Gamma}) = K(\mathbf{p}) + \Phi(\mathbf{q})$ is the Hamiltonian for N particles interacting via the WCA potential $\Phi(\mathbf{q}) = \sum_{i=1}^{N-1} \sum_{j>i}^{N} \phi(|\mathbf{q}_i - \mathbf{q}_j|)$ with $\phi(q) = 4[q^{-12} - q^{-6}]$, $q < 2^{1/6}$; $= 0$, otherwise. We assume the N particles populate a cubic cell of volume V in d Cartesian dimensions and that the system is periodic in the spatial coordinates $\mathbf{q}$.

We assume that the initial ensemble of phases characterised by a normalised N-particle phase space distribution $f(\boldsymbol{\Gamma}, 0) \sim \exp[-\beta H_0(\boldsymbol{\Gamma})]$, where $\beta$ is the usual Boltzmann factor $\beta = 1/k_B T$ and T is the absolute temperature. The dissipative flux, $J(\boldsymbol{\Gamma})$ is easily seen to be, $\dot{H}_0 \equiv -J(\boldsymbol{\Gamma})VF_c = -F_c \sum c_i \dot{y}_i$. Applying the master Fluctuation equation, (3), to this problem shows that

$$\ln\left[\frac{\Pr(-\beta \bar{J}_t V F_c = A)}{\Pr(-\beta \bar{J}_t V F_c = -A)}\right] = At. \tag{4}$$

It is important to note that the Boltzmann factor appearing in this equation refers to the temperature of the *initial* canonical ensemble. It does *not* refer to the time dependent temperature of the dissipative system. Because the colour field does work on the system, on average the system heats up as time increases. The entropy production inferred from linear irreversible thermodynamics [7] would be $\Sigma(t) = -J(t)VF_c / T(t)$ whereas equation (4) refers to the quantity defined in (2), $\Omega \equiv -\beta J(t)VF_c = -J(t)VF_c / k_B T(0)$. For convenience we will refer to $\Omega$ as the remnant entropy production.

The second system we consider is the same periodic set of interacting WCA particles, this time with no applied colour field. However, at $t = 0$, the system is at equilibrium under the influence of a nondissipative sinusoidal gravity field, $\Phi_g = \sum_{i=1}^{N} g \sin ky_i$, ( $k = 2\pi / V^{1/d}$ ). This field establishes an approximately sinusoidal density variation across the unit cell in the limit of low g. At $t = 0$, this field is removed and the system is monitored as it relaxes to equilibrium. As g becomes large, the particles become confined to the upper half the box if g is positive (see Figure 1). The initial distribution function for this system is chosen to be canonical:

$$f(\boldsymbol{\Gamma}, 0) \sim \exp\left(-\beta(K(\mathbf{p}) + \Phi(\mathbf{q}) + \Phi_g(\mathbf{q_y}))\right) \tag{5}$$

For t > 0, we observe the free relaxation of the density modulations towards equilibrium. It is straightforward, using equation (2), to find the the general dissipative function for this system:

$$t\overline{\Omega}_t \equiv \beta g \sum_{i=1}^{N} (\sin ky_i(t) - \sin ky_i(0)) \tag{6}$$

and from (3) the Fluctuation Theorem is:

$$\ln \frac{\Pr(\beta g \sum_{i=1}^{N} (\sin ky_i(t) - \sin ky_i(0)) = A)}{\Pr(\beta g \sum_{i=1}^{N} (\sin ky_i(t) - \sin ky_i(0)) = -A)} = A \tag{7}$$

In this equation $\beta$ again refers to the temperature of the initial ensemble rather than the time dependent temperature of the relaxing system. Equation (7) shows that as time increases it becomes overwhelmingly likely, that the initial density inhomogeneities will disappear. This particular proof of Le Chatelier's Principle can be generalised to arbitrary nondissipative perturbing fields, $\Phi_g$.

In order to test the Fluctuation Theorem for the two cases, we carried out molecular dynamics simulations of N = 32 WCA particles in two Cartesian dimensions. In both cases the initial system had a particle density of n = 0.4 and a temperature of T = 1.0. Molecular dynamics simulations using a Nosé-Hoover thermostat [8, 9] with a heat bath of of mass Q = 10 were used to sample the initial canonical ensembles for the two systems. Since the initial systems are ergodic, Monte-Carlo techniques could also have been used to generate the initial canonical distributions.

In the first case we measure the adiabatic response of the system to a colour field applied in the x-direction. From a single trajectory, initial canonically distributed phases $\mathbf{\Gamma}(0)$, were sampled at regular intervals. For these phases, a colour field was applied and the thermostat disabled and transient trajectories were generated according to the equations of motion:

$$\dot{\mathbf{q}}_i = \frac{\mathbf{p}_i}{m}$$

$$\dot{\mathbf{p}}_i = \mathbf{F}_i - \mathbf{i}c_i\,F_c \qquad\qquad\qquad (8)$$

where $\mathbf{F}_i = -\partial\Phi/\partial\mathbf{q}_i$, $F_c$ is the imposed colour field, $c_i = (-1)^i$ is the colour of the ith particle, which determines its response to the colour field. Simulations are carried out for systems employing a colour field of $F_c = 0.4$.

Figure 2 shows the distribution of the probability histogram for the time averaged value of the remnant entropy production, $\overline{\Omega}_t = -\beta\overline{J}_t VF_c$, for different trajectory segment lengths t = 0.2, 0.8, 1.6 and 4.0. As the trajectory segment becomes longer, the average value of the remnant entropy production increases due to the work done by the field, and the standard deviation of the distribution decreases due to the longer averaging time. In Figure 3, the value of ln[Pr(A)/Pr(-A)] is plotted as a function of A where $A = \overline{\Omega}_t$ or $A = \overline{\Sigma}\,/\,k_B = -\overline{\beta J}_t VF_c$. If the FT given by equation (4) is valid, a line of slope t will be obtained. Clearly, as expected, equation (4) is verified. However, if we also test the relationship using the irreversible thermodynamic entropy production rather than the remnant entropy production, a divergence from linear behaviour is observed.

In the second case we monitor relaxation of a system with an inhomogeneous density profile towards equilibrium. During this relaxation time, the equations of motion for the particles are simply Newtonian (no thermostats or external fields). The mean of the probability histogram for $\overline{\Omega}t = \beta g\sum_{i=1}^{N}(\sin ky_i(t) - \sin ky_i(0))$ shifts from zero when t approaches zero, to $-\beta g\langle\sin ky(0)\rangle$ as t→∞ since we expect that $\beta g\langle\sin ky(t)\rangle$ will approach zero as the density becomes homogeneous and the system approaches equilibrium. In this case, the variance of the distribution will approach a constant value as t increases. We test the FT for this system, given by equation (7) in Figure 4. The initial sinusoidal gravity field was set g=0.5 with trajectory segments of length t = 2.0 and t = 4.0. In both cases a straight line of slope unity is obtained, as expected from equation (7).

The FT given by equation (7) suggests that if $\overline{\Omega}t = \beta g \sum_{i=1}^{N} (\sin k y_i(t) - \sin k y_i(0))$ is positive, the logarithm of the probability ratio is positive and hence that $\overline{\Omega}t$ is more likely to be positive than negative. This means that the initial density fluctuation must decay from an inhomogeneous system where if $g > 0$, $\beta g \langle \sin k y \rangle < 0$ (see density profiles in Figure 1), to a homogenous system where $\beta g \langle \sin k y \rangle = 0$.

We have shown that phase space compression which results from the application of deterministic thermostats is *not* an essential element of systems which satisfy the Fluctuation Theorem. We have derived a generalisation of the Fluctuation Theorem which applies to an ensemble of adiabatic dissipative systems. The FT so derived shows that with increasing system size and observation time, it becomes exponentially likely that the dissipative flux will flow in the direction predicted by the Second Law of thermodynamics. However, unlike the situation for *thermostatted* nonequilibrium systems, the new adiabatic FT does not involve time averages of the entropy production that one would infer from standard irreversible thermodynamics. Further, we have verified that the corresponding FT which employs the standard expressions for the entropy production, is not valid.

We have also developed a version of the Fluctuation Theorem which applies to the free relaxation of isolated Hamiltonian systems towards equilibrium. This theorem constitutes a statistical mechanical proof of Le Chatelier's Principle, in the sense that as the observation time and system size increase, it becomes overwhelmingly likely that initial deviations from equilibrium will decay rather than grow spontaneously.

The Fluctuation Theorem would appear to run counter to Loschmidt's Umkehreinwand [10]. There are two points we make in this regard. Firstly, in our proof perfect time reversal symmetry is broken by the assumption of causality [2c]. We compute the required probabilities of phase space trajectory time averages, from *initial* rather than from *final* states. Few would object to this assumption. Had we computed the required probabilities from the final rather than the initial states then we would have derived an anti-Fluctuation Theorem [2c].

Secondly, since the equations of motion are time reversible, as Loschmidt observed, for every phase space trajectory its time reversed *antitrajectory* is also observable dynamically. The Fluctuation Theorem that we have derived gives an expression for the ratio of probabilities of observing time averaged properties from these time reversed trajectory pairs. Our computer simulation results confirm the validity of the theorem. The particular properties and ratio given by the Fluctuation Theorem confirm that in adiabatic dissipative systems, the sign of the entropy production is overwhelmingly likely to be in accord with the Second Law. In isolated Hamiltonian systems time evolution of initial inhomogeneities will with overwhelming likelihood be in accord with Le Chatelier's Principle and the system will in all likelihood, relax toward equilibrium.

**Acknowledgements**

We would like to thank the Australian Research Council for the support of this project. We would also like to thank Professor E.G.D. Cohen for helpful discussions.

**Figure Captions**

**Figure 1.** The density profile for systems of N = 32 particles in 2 Cartesian dimension at n = 0.4 and T = 1.0 with $\Phi_g = \sum_{i=1}^{N} g \sin k y_i$, where g = 0.5 (dashed line) or 10.0 (solid line) and $k = 2\pi / \sqrt{V}$.

**Figure 2.** Probability histograms for the time-averaged remnant entropy production obtained for a system of N = 32 particles in 2 Cartesian dimensions at n = 0.4 and T = 1.0 subject to a colour field of $F_c$ = 0.4 and averaged over trajectory segments of length t = 0.2, 0.8, 1.6 and 4.0.

**Figure 3.** The logarithm probability ratio of the remnant entropy production (x) and the irreversible thermodynamic entropy production (+) as a function of the time averaged remnant entropy production and the irreversible thermodynamic entropy production, respectively, for a system of N = 32 particles in 2 Cartesian dimensions at n = 0.4 and T = 1.0, subject to a colour field of $F_c$ = 0.4 and averaged over trajectory segments of length t = 0.8. If the FT given by (4) is valid, then the data will fall on a straight line of slope t (0.8), as shown by the solid line.

**Figure 4.** A test of the FT given by (7) for a system of N = 32 particles in 2 Cartesian dimension at n = 0.4 and T = 1.0, initially subject to a gravity field of g = 0.5. If (7) is valid a slope of unity is obtained, which is shown by the solid line. The trajectories were of length t = 2.0 (x) and t = 4.0 (+).

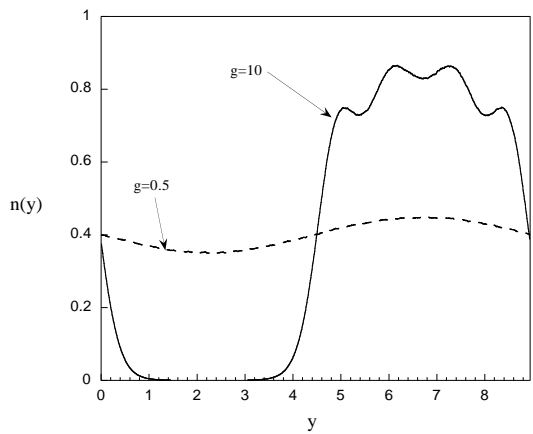

Figure 1

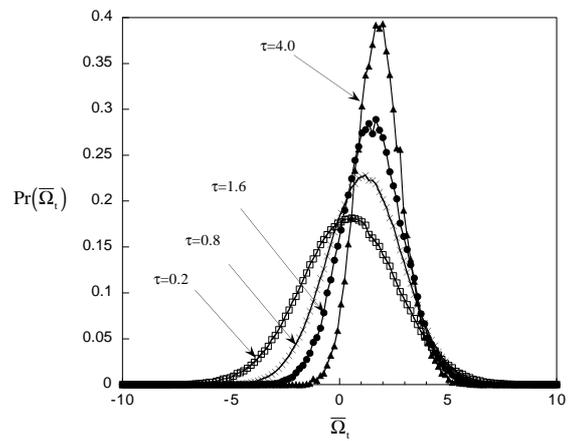

Figure 2

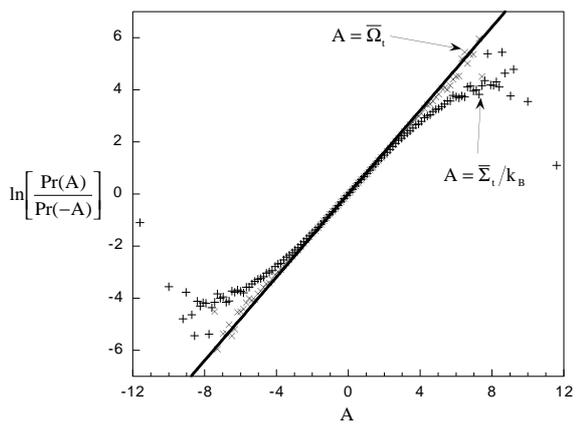

Figure 3

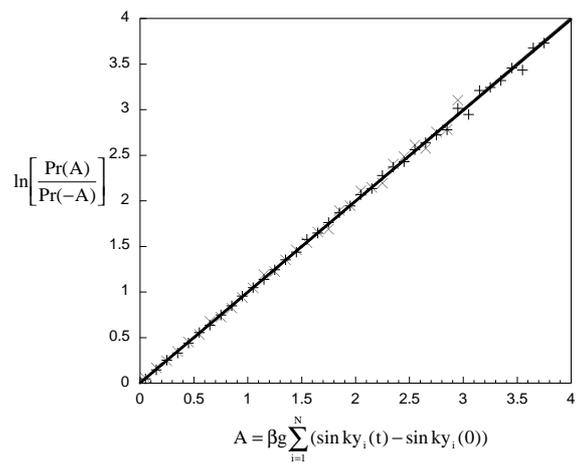

Figure 4